\documentclass{aa}             
\usepackage{amssymb,graphicx}  

\begin{document}

\thesaurus{06 (02.01.2 - 08.02.1 - 02.02.1 - 13.25.5 - 08.09.2 A0620-00)}

\title{Black hole soft X-ray transients: evolution of the cool disk and 
mass supply for the ADAF}
\author{E. Meyer-Hofmeister, F. Meyer}

\institute{Max-Planck-Institut f\"ur Astrophysik, Karl
Schwarzschildstr.~1, D-85740 Garching, Germany}

\offprints{emm@mpa-garching.mpg.de}

\date{Received:s / Accepted:}
\titlerunning {Black hole soft X-ray transients: disk evolution and ADAF}
\maketitle

\begin{abstract}
Using the black hole transient X-ray source  A0620-00 as an example we
study the physical interplay of three theoretical constituents for
modelling these transient sources: (1) the advection-dominated
accretion flow (ADAF) onto the central black hole, (2) the evaporation
of matter from the cool outer disk forming a coronal flow and (3)
standard disk evolution leading to outburst cycles by accretion disk
instability (dwarf nova mechanism).

We investigate the evolution of accretion disks during  quiescence
including the evaporation of gas in the
inner part of the disk. About 20\% of the matter is lost in
a wind from the corona. The mass flow rate
obtained from our model for the coronal flow towards the black hole,
is about half of the matter flowing over from the companion star. It
agrees with the rate independently derived from the ADAF spectral fits
by Narayan et al. (1997). About one third of the matter accumulates in
the outer cool disk. The computed disk evolution is consistent with the
observational data from the outburst in 1975. 
The evolution of the accretion disk until the instability is reached
shows that the critical surface density can not be reached for rates only
slightly less than the rate derived here for  A0620-00. Systems with
such accretion rates would be globally stable, suggesting that many
such faint permanently quiescent black hole X-ray binaries exist.


\end{abstract}

\section{Introduction}

Soft X-ray transients (SXT) are binaries in which a black hole or a
neutron star primary accretes matter from a main sequence or giant
secondary (van Paradijs \& McClintock 1995, Tanaka \& Shibazaki 1996). 
Several sources containing a black hole show outbursts every few
decades, which last a few
months. During the long-lasting quiescent state the systems
are very dim (McClintock \& Remillard 1990, McClintock et al. 1995). 
The X-ray luminosity observed during this state from binaries
containing a black hole is systematically lower than that from systems with
a neutron star. But the luminosity from the neutron star systems is
not as high as expected for accretion of all matter
at the neutron star's surface. For a recent compilation of ASCA
observations of SXTs in quiescence see Asai et al. (1998). It was
suggested that in the neutron star systems part of the matter is
thrown out by a magnetic propeller (Illarionov \& Sunyaev 1975) and
therefore the luminosity is reduced ( Asai et al. 1998, Menou et al. 1999b).

For the X-ray nova A0620-00 in quiescence the comparison of observed
optical and X-ray luminosity showed that the observed X-ray flux is
much less than expected if one assumes the
same accretion flow everywhere in the disk (see McClintock et al. 1995
and references therein). This observation led to the suggestion that
most of the thermal and kinetic energy is carried into the black
hole by an advection-dominated accretion flow (ADAF) and is not
radiated away. The ADAF model includes the
physics of the hot accretion flow (Narayan et al. 1996,
1997, for a recent review see Narayan et al. 1999) and is  
successful in describing the observed spectra of black hole SXTs in
quiescence. It provides strong evidence for the black hole nature
of the accreting primaries.

The configuration of the black hole SXT is the following. Matter from
the Roche-lobe filling secondary star flows over to an accretion
disk. In quiescence the disk is cool in the outer part. At a certain transition
radius, more accurately a transition region, the accretion flow is transformed
into a hot coronal flow and farther in no underlying cool disk exists
anymore (Esin et al. 1997). This coronal flow
continues inward and, at very high temperatures, becomes an
advection-dominated accretion flow. For a schematic drawing see Fig. 1.

\begin{figure}[t]
\includegraphics[width=8.8cm]{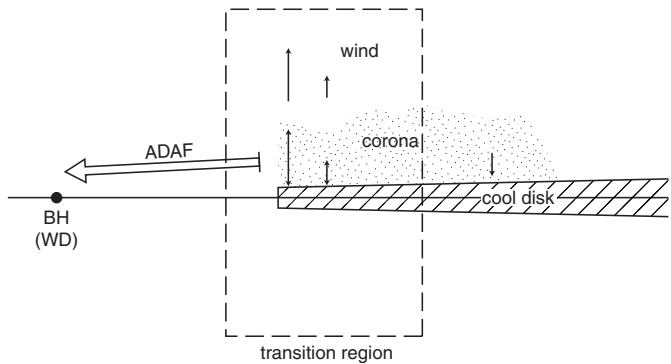}
\caption{Schematic drawing of cool disk, corona and ADAF}
\end{figure}

Our present investigation studies evaporation as the
source of the matter for the ADAF. We have shown earlier
(Meyer \& Meyer-Hofmeister 1994,  Liu et al 1995,
Liu et al. 1997) that evaporation of
matter from a cool disk into an overlying hot corona is important for the
evolution of accretion disks in binaries during 
quiescence. The same process works in dwarf nova disks and
in disks of X-ray novae (Mineshige et al. 1998). In dwarf nova
disks the inner part of the disk around the white dwarf primary can be
evaporated, depending on the amount of mass flow in the cool
disk. In black hole SXTs the horizon lies at much smaller radii and
the coronal flow continues into a probably more spherical
inner area filled with very hot gas. If the standard thin
accretion disk would extend to such small radii with the same mass flow
rate as in the outer regions the matter would become ionized and 
the disk instability would not allow the whole disk to remain in a quiescent
state. Lasota et al. (1996) pointed out that the inner part of the
thin disk has to be truncated.  

In Sect. 2 we discuss the physics of the evaporation process applied
to a black hole source. We point out that the interaction
of evaporation and mass accretion in the cool disk are similar for
dwarf novae and
transients. We compute the disk evolution using the well observed
system  A0620-00 as an example (Sect. 3). In Sect. 4 we compare our
results with
the findings from the spectral fit based on the ADAF model. In Sect. 5
we summarize the general features.

\section{Evaporation in disks around white dwarfs and black holes}
\subsection{The evaporation process in disks around black holes} 

We analyse the disks around black holes. The situation for disks
around neutron stars is more complex. Irradiation by the hot neutron
star surface has to be taken into account. But it is expected that for
both, black hole and neutron star sources the accretion flow near
the compact star occurs as an ADAF (Menou et al. 1999b).

The evaporation model decribes how above the cool inert disk a hot
self-sustained coronal layer is built up, fed by matter 
from the disk underneath (Meyer \& Meyer-Hofmeister 1994, Meyer 1999). The
vertical structure of the corona is established by the balance between
heat generation and
radiation losses at the coronal-chromospheric transition layer.
Based on a one-zone model Liu et al. (1995) computed the
evaporation rate ${\dot M_{\rm{ev}}}$ (evaporation into the corona above
and below the cool disk, both sides) for different white dwarf masses up to
$1.2\,M_\odot$. The rate varies with the  distance to the
primary star $r$ and its mass $M_1$ as

\begin{equation}
\dot M_{\rm{ev}}=10^{15.6}\,\,\left(\frac{r}{10^9 \rm cm}\right)^{-1.17}\,
\left(\frac{M_1}{M_\odot}\right)^{2.34} \,\,\,\,[\rm {gs^{-1}}].
\end{equation}

Due to wind loss the coronal mass flow
rate at the inner edge of the cool disk towards the compact object, 
${\dot M_{\rm{acc}}}$, is about 80\% of ${\dot M_{\rm{ev}}}$.

We checked the applicability of this result to larger compact object
masses with respect to the following physical conditions:
the electron-proton collision frequency is high enough that equilibrium is
established within the dynamical timescale, the mean free path of
transporting  electrons is small compared to the temperature scale
length in the conductive boundary layer. The ratio of advective to
conductive energy flux remains the same. All these three ratios remain
nearly invariant for large changes in central mass.

\subsection{The balance between mass accretion and evaporation.}
An important feature of the accretion disk in quiescence is the existence
of a transition between cool disk and hot coronal flow. To 
evaluate the transition radius $r_{\rm{trans}}$ one needs to know the mass
flow rate $\dot M_{\rm d}$ in the disk. The assumption of a
quasi-stationary disk allows such an estimate. To determine how this
location changes during the long period of quiescence one has to compute the
evolution of the disk. We describe here briefly earlier results to show the
similarity in the evolution of the inner disk in different binaries.
To follow such a disk evolution it is necessary to include conservation
of mass and angular momentum in the interacting cool disk and corona
(Liu et al. 1997). The angular momentum released farther inside by
the accreting matter flows outward in the corona and even forces a small
part of coronal matter to flow outward. This matter condenses
later in the outer cool disk (compare Fig. 4).

\begin{figure}[h]
\includegraphics[width=8.8cm]{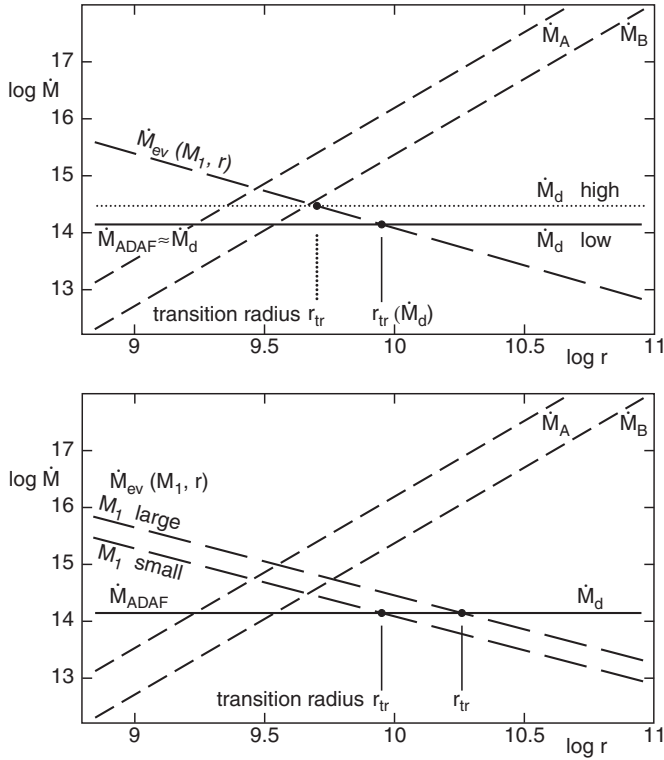}
\caption{Schematic drawing for the adjustment of the
transition radius $r_{\rm{tr}}$ (=location of the
transition between cool disk and coronal flow).
$r_{\rm{tr}}$ depends on the evaporation rate $\dot M_{\rm{ev}}$ 
and on the mass flow rate $\dot M_{\rm d}$ in the cool
disk, $M_1$ primary mass.
The upper panel shows $\dot M_{\rm{ev}}$ [Eq. 1] and $r_{\rm{tr}}$ for 2
 different rates $\dot M_{\rm d}$: for lower $\dot M_{\rm
 d}$ the transition occurs farther out. For
 explanation of the critical mass flow rates $\dot M_{\rm A}$ and $\dot M_{\rm
 B}$ see Fig. 3 where the corresponding values 
 $\Sigma_{\rm A}$ and $\Sigma_{\rm B}$ are given. The lower panel shows  
 $\dot M_{\rm{ev}}$ for 2 different primary masses and the resulting 
 change of $r_{\rm{tr}}$: for smaller $M_1$ the transition
 occurs farther in. (cgs units)}
\end{figure}

\subsubsection{Ordinary dwarf nova}
Liu et al. (1997) studied the evolution of the disk around a
$1\,M_\odot$ primary star in an ordinary dwarf nova system 
with a relatively high mass overflow rate of $\dot M$=
$2\,10^{-9}M_\odot/ \rm {yr}$ from the secondary. Disk evaporation 
creates a hole in the inner disk and the transition remains near
$r=4\,10^{9}$cm while matter piles up in the outer disk until the onset
of instability occurs near $r=10^{10}$cm. Without evaporation this
would have happened near to the white dwarf surface. For a comparison of
accretion disk instabilities in dwarf novae and transient X-ray
sources see Cannizzo (1998).

\subsubsection{WZ Sge stars}
WZ Sge stars, a subgroup of dwarf nova (Ritter \&Kolb 1998), have also
very long outburst recurrence times similar to black hole SXTs considered
here (Kuulkers 1999). This makes WZ Sge an interesting candidate to
study the effect of evaporation.
For the parameters we had chosen according to observations
(Liu et al. 1998) we found that during the first years of
quiescence the mass flow rate in the
inner disk is lower than the evaporation rate,
 $\dot M_{\rm d}  \le \dot M_{\rm{\rm{ev}}}$ ,
and a hole is formed. With the increasing accumulation of matter 
${\dot M_{\rm d}}$ rises and the transition shifts towards the white
dwarf until after about 9 years the hole is closed. Evaporation goes
on all the time, even after the inner edge of the disk has reached the
white dwarf surface, until the outburst is triggered (Fig. 1,
Meyer-Hofmeister et al. 1998). 

\subsubsection{General aspects}
Generally the inner edge of the cool disk will be established where
the evaporation rate equals the mass flow rate in the disk. This
balance determines the transition radius if the mass of the primary
and the mass flow rate in the disk are known. The fact that the disks are
quasi-stationary during long quiescence intervals allows a first estimate
for $r_{\rm{trans}}$. We show in a schematic drawing how
$r_{\rm{trans}}$
changes if 
${\dot M_{\rm d}}$ changes (Fig. 2a) and how a different primary mass
influences the evaporation rate $\dot M_{\rm{ev}}$ and therefore the location
of the transition.

\section{The evolution of the accretion disk in the black hole soft
X-ray transient A0620-00 during  quiescence}

The system A0620-00 (Nova Mon 1917,1975), discovered by {\it{Ariel}} in
1975 (Elvis et al. 1975), is one of the best studied X-ray novae.
For a recent detailed compilation of the observations for  A0620-00
see Kuulkers (1998). The very first theoretical investigations of the
accretion disk ( Huang \& Wheeler 1989 and Mineshige \& Wheeler 1989)
aimed to clarify whether the outburst behaviour of A0620-00 could be
understood as an accretion disk instability. In later work the
emphasis was put mostly on modelling the observed exponential decay of
the outburst luminosity. Recently Cannizzo (1998) reviewed these
investigations and included an example of outburst modelling based on
disk evolution with an assumed law for loss of matter from the cool
inner disk, but without the flow of angular momentum in the combined
disk + coronal accretion.

Our investigation concentrates on the evolution of the accretion disk
during quiescence using a computer code which includes the physics of
interaction of disk and corona (see Sect. 3.1). Observations for this
state are more difficult, but have
been carried out in the optical, UV, EUV  and X-ray bands. For a
description of the observations and deduction of system parameters
see Narayan et al. (1996, 1997). The orbital period is 7.8\,hr
(McClintock et al. 1983).
The mass of the black hole follows from the mass function if the
inclination is known. 
Narayan at al. (1996,1997) considered different values for the
inclination i and the corresponding primary mass $M_1$: i=$70^{\circ}$
and $M_1$=$4.4\,M_\odot$, i=$40^{\circ}$ and $M_1$=$12\,M_\odot$ and
in the later paper also i=$55^{\circ}$ and $M_1$=$6.1\,M_\odot$, as
recently suggested by Barret et al. (1996). 

\subsection {Technique of computations}
For our modelling of the disk evolution we take BH masses of
$4\,M_\odot$ and $6\,M_\odot$.
The viscosity parameter (Shakura \& Sunyaev 1973) assumed was
$\alpha_{\rm{cool}}$= 0.05, a standard value for the cool disk in dwarf
nova outburst modelling. 
We assume that only a small amount of matter was left
over in the disk after the last outburst (the value of the critical
surface density $\Sigma_A$, compare Fig. 3, corresponds to
$\alpha_{\rm{hot}}$=0.3. Note that the viscosity in the hot state used in
the ADAF model is not relevant for the modelling of the quiescent
phase). We start with an outer disk radius of
9.2\,10$^{10}$cm (geometry connected with the orbital
period, secondary mass taken as 0.4$M_\odot$). During the evolution
the disk grows to 10$^{11}$cm.
The same code was used as for WZ Sge (solution of the diffusion
equation, changes of outer radius due to conservation of mass and
angular momentum, evaporation in the inner disk into the coronal flow,
changes of the inner edge of the disk, Meyer-Hofmeister et al. (1998)).

\subsection {Constraints for the disk evolution}
The lightcurve of A620-00 in B magnitude for the outburst in 1917, 
reconstructed by Eachus et al. (1976), has a shape similar to the lightcurve
of the 1975 outburst (Lloyd et al. 1977), indicating that 
the two outbursts were similar. Though it is not certain
that no outburst was missed it is generally assumed that no outburst
ocurred in between.
Basic constraints on the quiescent disk evolution are then 
the two facts: (1) the outburst recurrence time of 58 years 
and (2) the amount of matter stored in the disk and released in
the outburst. The amount of energy in the 1975 outburst was estimated
as 3-4\,$10^{44}$erg (McClintock et al. 1983, White et al. 1984).
This estimate refered to a neutron star primary and to
isotropic radiation, a distance of 0.9 kpc was assumed. If we use 1 kpc for
the distance as Narayan et al. (1997) and assume a black hole and the
disk seen under an inclination of
$70^\circ$  we obtain for the matter accreted in the outburst
${\Delta} M$= 6-8\, $10^{24}$ g (for the inclination see also the
newer analysis of Barret et al. 1996).

\vspace* {1.5cm}

\begin{figure}[h]
\includegraphics[width=8.8cm]{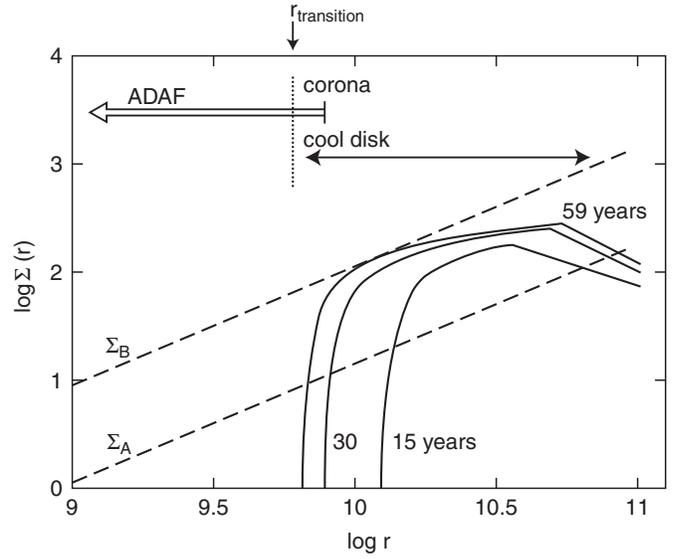}
\caption{ 
Disk evolution of A0620, ''standard case'', accumulation of matter in
quiescence.
$\Sigma_{\rm A}$ and $\Sigma_{\rm B}$ critical surface densities. For
the surface densities between these values 
two states, hot and cold, are possible. When accumulation of matter reaches
${\Sigma_{\rm B}}$ an outburst sets in}
\end{figure}

The total amount of matter is closely related to the rate of mass overflow
from the companion star. We study the situation for rates 
in the range ${\dot M}=10^{-10}M_\odot/\rm{yr}$ (McClintock et
al. 1995) to $2\,10^{-10}M_\odot/\rm{yr}$, the latter corresponding to the
results of Narayan et al. (1997).
Narayan et al. (1996, 1997) studied the ADAF in A0620-00 in two
investigations.
The first results from the fit to the spectra were mass flow
rates in the disk (assuming that the same rate flows through
the outer disk and in the ADAF) of only 7\,$10^{-12}$ to
2\,$10^{-11}M_\odot/\rm{yr}$ for a primary mass of 4.4\,$M_\odot$. But in
the second investigation, based on very
elaborate calculation techniques and testing different heating of
electrons, the resulting mass flow rates in the ADAF for most of the models 
lie in the range 1 to 
$1.5\,10^{-10}M_\odot/\rm{yr}$. Most of the models were computed
for a primary mass of 6.1 $M_\odot$, but the resulting mass flow rate
was about the same for 4.4 $M_\odot$ as taken in the
earlier investigation. It was found that nearly all the optical flux
comes from the ADAF, the outer disk being quite negligible. From our
point of view this means that the mass overflow rate is higher by the
additional amount of matter accumulated in the outer disk for the outburst and
by the amount lost in the wind when the coronal flow is formed. Our
earlier investigations of the
evaporation process gave about 20\% wind loss (Liu et al. 1995).
The mass overflow rate therefore might be as high as
$2\,10^{-10}M_\odot/\rm{yr}$. In an earlier investigation
(Meyer-Hofmeister \& Meyer 1999) first results for an evolution based
on a low mass overflow rate were presented.

\begin{figure}[hb]
\includegraphics[width=8.8cm]{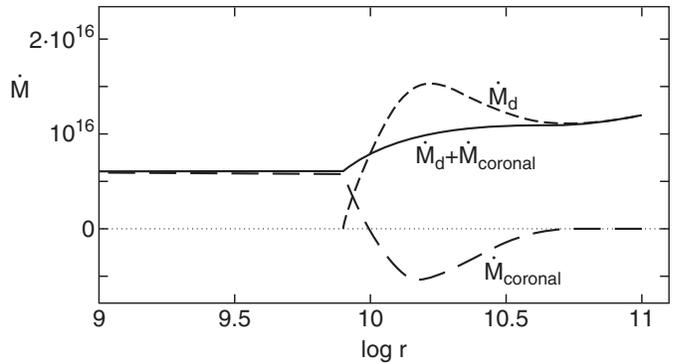}
\caption{ 
Disk evolution of A0620,  ''standard case', with $\dot M_{\rm{overflow}}=
1.9\,10^{-10}M_\odot/{\rm{yr}}$, mass flow rates 30 years after outburst (
compare Fig. 3): $\dot M_{\rm d}$ mass flow in the
standard cool disk, $\dot M_{\rm{coronal}}$ hot coronal flow and total
flow. The negative values for the coronal flow indicate a flow outward
(for conservation of angular momentum during disk evolution).} 
\end{figure}

\subsection{Results for disk evolution, standard case} 
The aim of our computation is to find out whether one can model the disk
evolution so that it complies with
the constraints on recurrence time and amount of matter stored in the
disk $\Delta M$and also agrees with mass flow rate in the ADAF and transition
radius derived from the spectral fit by Narayan et al. (1997). Our
``standard case' represents such a consistent disk evolution. 
We take the following parameters: primary
mass $M_1$=4 $M_\odot$, mass overflow rate
1.9\,$10^{-10}M_\odot/\rm{yr}$. In Fig. 3 we show the evolution of the thin
cool accretion disk with a hot corona above. The onset of the
outburst occurs after 59 years. During quiescence
the surface density in the disk increases steadily everywhere as the
distributions after 15, 30 and 59 years show. The
surface density in the outer disk determines the total amount of
matter stored at the onset of the outburst $\Delta M$, in our example
6.9\, $10^{24}$g.
The transition radius adjusts so that
the evaporation rate is the same as the mass flow
rate in the inner cool disk (if the transition would occur only in a
very narrow radial interval, the evaporation rate and the mass flow
rate there would be the same, but in our computations the evaporation
of matter into the hot corona is spread out over a transition area).
In our standard case we get for the location of the inner edge of the cool disk
$r_{\rm{trans}}= 6.8\,10^9$cm. 
In Fig. 4 we show the mass flow in the disk after
30 years of quiescence for the ``standard case', a situation comparable to the present state of
A0620. 6\,$10^{15}$g/s, half of the overflow rate,
flows inward towards the black hole. In the transition region a fraction
of the matter flows outward in the corona and condenses again in the
cool disk. Such an outward flow is necessary for the conservation of
angular momentum in the co-existing cool disk and coronal disk. In the
transition region matter is also lost via a wind from the corona. 
This division of mass flows is about the same over the total computed
quiescence evolution.

\subsection{Results for disk evolution, variation of mass overflow
rate and viscosity}
We also studied the disk evolution for other parameters.
The smaller the rate $\dot M_{\rm{overflow}}$ the more
time is needed to accumulate enough matter to reach the critical
surface density $\Sigma_{\rm B}$. The balance of mass flow in the disk and
evaporation determines the transition radius. A lower rate
$\dot M_{\rm{overflow}}$ results in a more extended hole (compare
Fig. 2, upper panel) and the critical
surface density can then only be reached farther out in the disk. This
means the surface density is higher everywhere in the disk and
therefore more matter piles up in the outer disk. The quiescence lasts
longer. We found that for an overflow rate of
1.8\,$10^{-10}M_\odot/\rm{yr}$, only a little lower than in our standard case,
no outburst occurs anymore and the disk is stationary. Such a system would not
easily be recognized. On the other hand a slightly higher rate
triggers the outburst early (see Table 1). For 
a slightly higher viscosity parameter $\alpha_{\rm cool}$=0.06 the
accumulated amount of matter $\Delta M$ is less and the recurrence time is
shorter than for the standard value 0.05. (Small variation of the
parameters within the allowed range would lead to exact agreement with
the observed recurrence time).

\begin{table}
\caption{Parameters of models of A0620}
\begin{tabular}{@{}
  c@{\hspace{4pt}}|@{\hspace{6pt}}
  c@{\hspace{4pt}}
  c@{\hspace{4pt}}
  c@{\hspace{4pt}}
  c@{\hspace{4pt}}
  c@{\hspace{4pt}}r@{}}
\hline
& & & & & & \\
model & $M_1$ & $\alpha_{\rm cool}$ &${\dot M}_{\rm overflow}$& recurrence &
$\Delta M$ & $r_{\rm trans}$ \\
& & & & time & & \\& $[M_{\odot}]$ & &[$10^{-10}\frac{M_{\odot}}{\rm
yr}$] & [\rm yr] & [$10^{24}\rm g$] &
[$10^9\rm cm]$\\
\hline
& & & & & & \\
1a & & & 1.8 & stationary & 6.9 & 7.6 \\
1b & 4 & 0.05 & 1.9 & 59 & 6.9 & 6.8  \\
1c & & & 2.0 & 44& 6.9 & 6.5 \\
& & & & & & \\
\hline
& & & & & & \\
2a & & & 1.8 & stationary & 5.7 & 7.6 \\
2b & 4 & 0.06 & 1.9 & 50 & 5.8 & 7.0  \\
2c & & & 2.0 & 38& 5.9 & 6.6 \\
& & & & & & \\
\hline
& & & & & & \\
3a & & & 3.0 & stationary & 12.3 & 10.7 \\
3b & 6 & 0.05 & 3.2 & 112 & 13.1 & 9.7  \\
3c & & & 3.4 & 55& 13.1 & 9.9 \\
& & & & & & \\
\hline
\end{tabular}
\end{table}

\begin{figure}[ht]
\includegraphics[width=8.8cm]{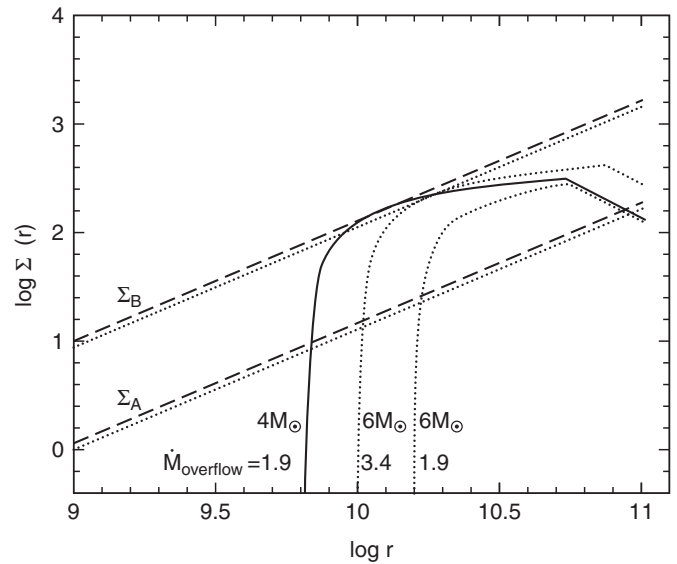}
\caption{ 
Comparison of accumulation of matter during quiescence for primary
mass 4 and 6$M_\odot$, mass overflow rates in $10^{-10}M_\odot/\rm{yr}$ :
solid line surface density distribution at the
onset of outburst (same as in Fig. 3), dotted lines onset of outburst
for a 6$M_\odot$ black hole: due to more efficient evaporation the
hole is larger and a higher mass overflow rate $3.4\,10^{-10}M_\odot/\rm{yr}$
instead of $1.9\,10^{-10}M _\odot/\rm{yr}$ is needed to produce an
outburst (in our example after 55 years). The low rate of
$1.9\,10^{-10}M_\odot/\rm{yr}$ would lead to a stationary disk without
an outburst.} 
\end{figure}

\subsection {Results for disk evolution, more massive primary star}

For higher primary mass the evaporation is more efficient (compare
Fig. 2, lower panel). The transition radius is larger. A higher mass
overflow rate is necessary to compensate for this. Assuming a black hole of
6\,$M_\odot$ and a rate of $3.4\,10^{-10}M_\odot/\rm{yr}$ we get an outburst
after 55 years. Almost twice as much matter is stored in the disk
compared to the ``standard case'' with the 4\,$M_\odot$ black hole. 
In Fig. 5 we show the
distribution of surface density at the onset of the outburst for 4 and
6\,$M_\odot$ together with the slightly different critical surface
densities for these primary masses. The rate adequate for the
triggering of the outburst for 4\,$M_\odot$ would  only give a
stationary disk for 6\,$M_
\odot$, far from an instability. For an even more massive primary of
12\,$M_\odot$ a very high mass overflow rate would be necessary to
reach an outburst. This would not agree with the constraints for
A0620, but might be the case for other black hole systems.

\section {Comparison of the results from disk evolution with the
results from the ADAF model}

\begin{table*}
\caption{Comparison of the results from disk evolution with the
transition radius from the $H_\alpha$ emission lines and the mass
accretion rate in the ADAF.}
\begin{center}
\begin{tabular}{
  l@{\hspace{20pt}}
  r@{\hspace{20pt}}
  r@{\hspace{20pt}}}
\hline
& & \\
$M_1$ & 4.4$M_\odot$ & 6.1$M_\odot$ \\
& & \\
log $r_{\rm{trans}}/{\rm{cm}}\,\, (H_\alpha)$ & 10.12 &10.06 \\
& & \\
${\dot{M}_{ADAF}}$* & 1.45$\times10^{-10} \frac {M_\odot}{\rm yr}$ 
& 1.31$\times 10^{-10} \frac {M_\odot}{\rm yr}$\\
& & \\
\hline
& & \\
$M_1$ & 4$M_\odot$ & 6$M_\odot$ \\
 & & \\
log $r_{\rm{trans}}/{\rm{cm}}$ (disk evolution **) & 10.0 to 9.9
 & 10.2 to 10.1
 \\
& & \\
$\dot M_{\rm{coronal}}$ (disk evolution **) & 0.8 to 1.1$\times
10^{-10} \frac {M_\odot}{\rm yr}$ & 1.2 to 1.6$\times
10^{-10} \frac {M_\odot}{\rm yr}$ \\
& & \\
\hline
\end{tabular}
\end{center}

$^*$ models 1 and 3 of Narayan et al. (1997)\\
$^{**}$ models 1b and 3b (c.f. Table 1), values 20 to 30 years after outburst

\end{table*}

In Table 2 we summarize the comparison of the results from our disk
evolution computations and the results from the spectral fit of the ADAF
model to the observations.
The transition radius $r_{\rm{trans}}(\rm {H_\alpha)}$ used for the ADAF fit
is estimated from observations on the basis of the largest velocity
$v_{\rm{max}}$ seen in the ${\rm H}_\alpha$ line,
2100\,$\rm{kms^{-1}}$ (Marsh et al. 1994, Orosz et al. 1994). For the
inclinations $70^\circ$ and $55^\circ$ and the related black hole masses 
the values $r_{\rm{trans}}$ follow, the smaller value for the smaller
inclination and higher primary mass. The ADAF fit yields a mass
flow rate $\dot M_{\rm {ADAF}}$. This result does not depend sensitively on
$r_{\rm{trans}}$. 
Our computations are carried out for two values of viscosity,
$\alpha_{\rm cool}$=0.05, the standard value from dwarf nova outburst
modelling and 0.06. We take 4 and 6$M_\odot$ for the primary mass.
We consider only cases of disk evolution which comply with the constraints
(1) recurrence time of 59 years of A0620 and (2) amount of
matter accumulated for the outburst in agreement with the outburst
energy (only a rough value). For 6$M_\odot$ the second constraint is
less well met. The flux measurements used for the ADAF fit 
were obtained many years apart as pointed out by Narayan et
al. (1996). For our models 1b and 3b (compare Table 1) we therefore give
the transition radius and the coronal mass flow rate (as supply for the
ADAF) for the evolutionary times of 20 and 30 years after the outburst.
This documents the change of values during evolution. 
Comparing the rates $\dot M_{\rm{ADAF}}$ and $\dot M_{\rm{coronal}}$ the values
are closer for the 6$M_\odot$ primary. But we have mentioned already
that in this case somewhat more matter is accumulated than estimated from the
outburst energy. Keeping in mind that here the results are brought
together from very
different physical processes, (a) the radiation from the innermost very
hot sphere and (b) the evolution of the cool disk with evaporation of matter
into a coronal layer, the derived rates $\dot
M_{\rm{ADAF}}$ and $\dot M_{\rm{coronal}}$ are remarkably similar.

\section {General features of the ADAF formation and conclusions}  
The computation of the disk evolution for a set of parameters allows us
to learn about the general features of the transition of the cool disk
to a coronal disk.
\subsection {Location of the transition}
As pointed out in Sect. 2 the transition radius is determined by
the balance of mass flow in the disk and evaporation. This situation
is the same in disks around white dwarfs and around black holes. In
dwarf nova systems for a high mass flow rate, the transition radius
might move in to the white dwarf surface or even disappear. Then no hole is
created in the disk but evaporation still goes on continuously. In disks
around black holes, as for example in A0620, for the low accretion
rates in quiescence a transition at a distance of about $10^{10}$cm
from the black hole results.

The transition radius can be estimated without following the disk
evolution if one assumes a mass flow rate in the disk. But the
evaluation of radii together with the disk evolution includes the
constraints on the evolution and is therefore more firm. Taking the
transition radius and the rate $\dot M_{\rm{ADAF}}$ and calculating the
evaporation rate one can compare these results. This was done in the
investigations of Meyer (1999) and also by Liu et al. (1999). In
the latter work this comparison was also performed for
advection-dominated accretion flows onto massive black holes in
galactic nuclei.

\begin{figure}[ht]
\includegraphics[width=8.8cm]{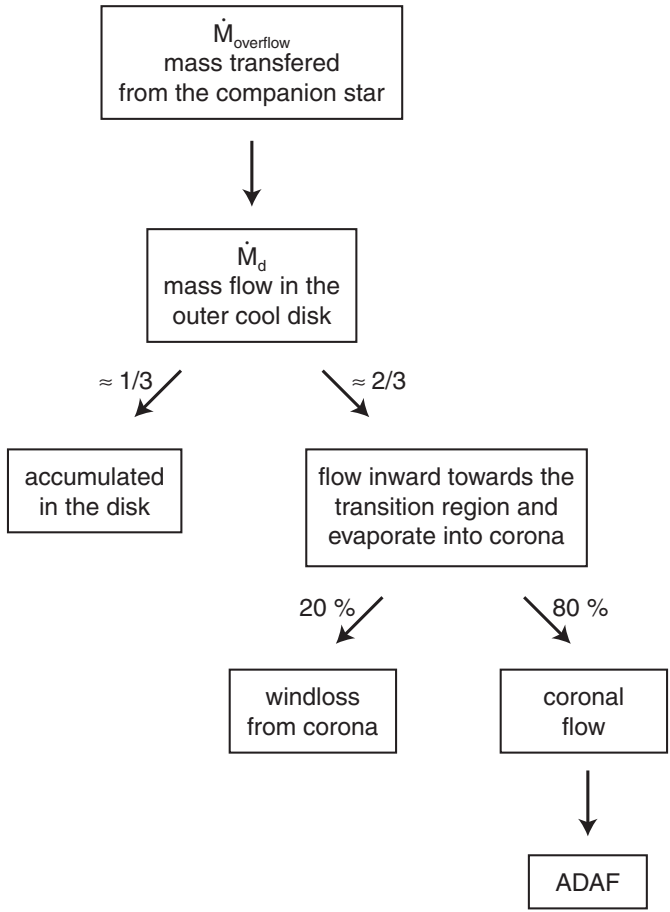}
\caption{ Rate of the advection dominated accretion flow resulting
from mass transfer from the companion star, model for A0620} 
\end{figure}

\subsection {The relation between the mass overflow rate and the mass
flow rate in the corona}
For the ADAF fit it was assumed that the mass flow rate in the outer cool
disk and in the inner very hot spherical region would be the
same. Since Narayan et al. (1997) found that most of the optical light
also originates in the ADAF the mass flow rate in the outer disk is
not important for the spectrum. But it is of interest to find out what
the mass overflow rate from the companion star is; in our best example
(model 1b) it is $1.9\,10^{-10}M _\odot/\rm{yr}$. 
This raises interesting questions about the nature of the
magnetic braking mechanism in this binary (for a discussion see King
et al. 1996 and Menou et al. 1999a).

From our computations we find that, rather independent of the time
passed after the last outburst, about 1/3 of the matter flowing over
from the secondary is stored in the outer disk for the next outburst.
The other 2/3 is evaporated into the corona. From this amount 20 \%
is lost in a wind and the remaining 80\% flows inward in the corona
and provides the matter for the ADAF in the very hot inner region close
to the black hole. This means that about half of the matter flowing over
from the companion star accretes onto the black hole in quiescence.
In Fig. 6 we show this division of the flow of matter in the cool
standard disk and the corona.

\subsection {Outbursts of X-ray transients} 
Our results for disk evolution show (compare Table 1) that small
changes in the mass overflow rate from the companion star affect
the outburst recurrence time essentially. For rates only slightly
less than the rate which provides the onset of instability after 59 years
(example A0620-00) the critical surface density can not be reached,
the system is globally stable. Probably many such systems exist,
faint persistent soft X-ray binaries.

\end{document}